\documentclass[aps,prl,twocolumn,preprintnumbers,superscriptaddress,nofootinbib,floatfix]{revtex4-1}
\usepackage[colorlinks=true,pdfstartview=FitV,breaklinks=true]{hyperref}

\usepackage[utf8]{inputenc}
\usepackage{graphicx}
\usepackage{booktabs}
\usepackage{tabularx}
\usepackage{textcomp}
\usepackage{xspace}
\usepackage{amsmath}
\usepackage{amsfonts}
\usepackage{amssymb}
\usepackage{comment}
\usepackage[dvipsnames,table]{xcolor}

\hypersetup{urlcolor=BlueViolet,
	    citecolor=Plum,
	    linkcolor=PineGreen}
\usepackage[official]{eurosym}
\definecolor{lightgray}{rgb}{0.9,0.9,0.9}	    
\definecolor{green}{rgb}{0,0.5,0}
\definecolor{red}{rgb}{1,0,0}
\definecolor{blue}{rgb}{0,0,0.5}

\newcommand\codo[1]{{\tt #1}}

\begin{document}

\title{Axion minicluster streams in the Solar neighbourhood}

\author{Ciaran A. J. O'Hare}\email{ciaran.ohare@sydney.edu.au}
\affiliation{School of Physics, Physics Road, The University of Sydney, NSW 2006 Camperdown, Sydney, Australia}

\author{Giovanni Pierobon}\email{g.pierobon@unsw.edu.au}
\affiliation{School of Physics, The University of New South Wales, NSW 2052 Kensington, Sydney, Australia}

\author{Javier Redondo}\email{jredondo@unizar.es}
\affiliation{CAPA \& Departamento de Fisica Teorica, Universidad de Zaragoza, 50009 Zaragoza, Spain}
\affiliation{Max-Planck-Institut f\"ur Physik (Werner-Heisenberg-Institut), F\"ohringer Ring 6, 80805 M\"unchen, Germany}

\smallskip
\begin{abstract}
A consequence of QCD axion dark matter being born after inflation is the emergence of small-scale substructures known as miniclusters. Although miniclusters merge to form minihalos, this intrinsic granularity is expected to remain imprinted on small scales in our galaxy, leading to potentially damaging consequences for the campaign to detect axions directly on Earth. This picture, however, is modified when one takes into account the fact that encounters with stars will tidally strip mass from the miniclusters, creating pc-long tidal streams that act to refill the dark matter distribution. Here we ask whether or not this stripping rescues experimental prospects from the worst-case scenario in which the majority of axions remain bound up in unobservably small miniclusters. We find that the density sampled by terrestrial experiment on mpc-scales will be, on average, around 70--90\% of the average local DM density, and at a typical point in the solar neighbourhood, we expect most of the dark matter to be comprised of debris from $\mathcal{O}(10^2$--$10^3)$ overlapping streams. If haloscopes can measure the axion signal with high-enough frequency resolution, then these streams are revealed in the form of an intrinsically spiky lineshape, in stark contrast with the standard assumption of a smooth, featureless Maxwellian distribution---a unique prediction that constitutes a way for experiments to distinguish between pre and post-inflationary axion cosmologies. 
\end{abstract}

\maketitle

\begin{figure*}
    \centering
    \includegraphics[width=0.99\textwidth, trim = 35mm 0mm 0mm 0mm, clip]{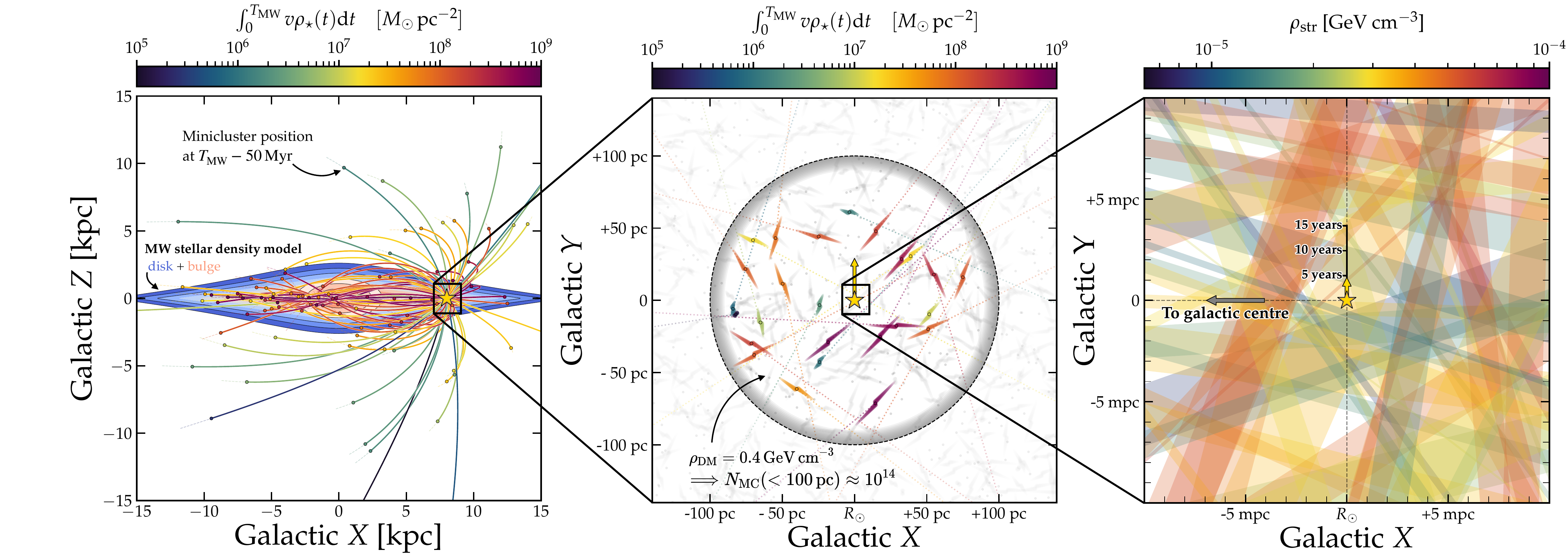}
    \caption{Schematic of our study. {\bf Left}: We begin by modelling the orbits of many miniclusters all ending at the Solar position today, $(X,Y,Z) \approx (8,0,0)$~kpc. For each orbit, we draw random encounters with stars from a MW stellar density model that includes the thin and thick disks, and the bulge. We colour the minicluster orbits by the integral of this density along the orbit. For clarity, we display only the last 50 Myr of the orbit. {\bf Centre}: Zooming in on a 100-pc-radius sphere around the Sun, we illustrate how the axions stripped from their host minicluster are elongated into tidal streams of $\mathcal{O}({\rm pc})$ length. {\bf Right}: Zoom in on $\sim$mpc scales relevant for experiments where the network of tidal streams sums to give the local density in axions.}
    \label{fig:main}
\end{figure*}

\noindent \textbf{\textit{Introduction}}.---Axions~\cite{Peccei:1977hh, Peccei:1977ur, Weinberg:1977ma, Wilczek:1977pj, Kim:2008hd, Kim:1979if, Shifman:1979if, Dine:1981rt, Zhitnitsky:1980tq} are one of the most popular explanations for the makeup of galactic dark matter (DM) halos (for reviews, see e.g~\cite{DiLuzio:2020wdo, Chadha-Day:2021szb, Irastorza:2018dyq, Semertzidis:2021rxs, Adams:2022pbo,OHare:2024nmr}). There are several well-motivated cosmological production scenarios for axion DM, but one of the most interesting and predictive examples is when the symmetry-breaking phase transition that births the axion occurs \textit{after} inflation. Dedicated numerical simulations studying this scenario have flourished in recent years~\cite{Fleury:2015aca, Klaer:2017ond, Vaquero:2018tib, Gorghetto:2018myk, Buschmann:2019icd, Gorghetto:2020qws, Buschmann:2021sdq, OHare:2021zrq}. 

It has been known for many years~\cite{Hogan:1988mp, Kolb:1994fi, Kolb:1995bu} that the complicated multi-scale dynamics of the axion field that necessitate the constriction of these simulations, also implies that the DM distribution in this scenario will inherit inhomogeneities on scales set by the horizon at the QCD phase transition. This results in the majority of DM becoming bound inside planetary-mass structures called \textit{miniclusters}~\cite{Hogan:1988mp, Kolb:1994fi, Kolb:1995bu, Zurek:2006sy, Kolb:1994fi, Hardy:2016mns, Davidson:2016uok, Enander:2017ogx, Blinov:2019jqc,Dandoy:2023zbi,Fairbairn:2017dmf,Fairbairn:2017sil, Croon:2020wpr, Ellis:2022grh,Edwards:2020afl}. To put it plainly: miniclusters would be too sparsely distributed for us to have a realistic chance of encountering one, so prospects for detecting DM axions in the lab rest upon whether or not miniclusters survive in our galaxy today.

In this letter, we address the detectability of DM axions by quantifying their distribution in the solar neighbourhood. Broadly speaking, DM in this scenario can be thought of in terms of three distinct populations. Firstly, there are the axions that never end up inside miniclusters to begin with, existing instead in the ``minivoids'' between them~\cite{Eggemeier:2022hqa}. Secondly, there are the miniclusters themselves, which lock up more than 80\% of the mass of DM prior to galaxy formation~\cite{Eggemeier:2019khm}. Lastly, there is the minicluster debris---axions tidally stripped from their hosts as they orbit the Milky Way (MW)~\cite{Tinyakov:2015cgg, Dokuchaev, Kavanagh:2020gcy,DSouza:2024flu,Shen:2022ltx}. 
The worst-case scenario for direct detection experiments (referred to as ``haloscopes'') is if we consider the axions only from the minivoids~\cite{Eggemeier:2022hqa}. However, tidal disruption will act to spread the DM across a wider volume~\cite{Tinyakov:2015cgg}, so observational prospects could be rescued if we account for this.

To that end, we have performed Monte-Carlo simulations that begin with a realistic population of miniclusters taken from early-Universe simulations, and are then evolved as they orbit the MW. For each minicluster, we calculate how much mass is stripped from it, and over what length scale this mass is spread so that we can build a model for the resulting stream network. Figure~\ref{fig:main} shows an illustration of the stages of this calculation. We then use these results to create example haloscope signals, as in Fig.~\ref{fig:signal}.

\noindent \textbf{\textit{Initial miniclusters}}.---To start our calculation, we need to know the initial distribution of minicluster masses and density profiles. We source this information from the most recent N-body simulations of minicluster formation, which themselves begin from initial conditions left over from lattice simulations of axions around the QCD phase transition~\cite{Pierobon:2023ozb}. The most pertinent result of these simulations is the two types of miniclusters that form---what we refer to as \textit{merged} miniclusters, and \textit{isolated} miniclusters. The former result from hierarchical merging and develop Navarro-Frenk-White (NFW) density profiles~\cite{Eggemeier:2019jsu, Xiao:2021nkb, Pierobon:2023ozb}, while the latter form from the prompt collapse of isolated overdensities and then do not undergo many substantial mergers---retaining the power-law profiles usually associated with self-similar collapse, $\rho \propto r^{-\tilde{\alpha}}$~\cite{Zurek:2006sy}. In our baseline set of results, we adopt $\tilde{\alpha} = 2.71$ taken from fitting their density profiles at the latest times in our simulation~\cite{Pierobon:2023ozb}.

Isolated miniclusters are the most abundant, however, they collectively make up very little of the total DM since they have masses $M\lesssim 10^{-12}\,M_\odot$. Merged miniclusters, on the other hand, have masses $M\gtrsim 10^{-12}\,M_\odot$, so comprise most of the DM by mass. We draw minicluster masses from a broken power-law mass function representing these two populations: $\textrm{d}n/\textrm{d}\log M\propto M^\gamma$, with $\gamma = -0.8$ for isolated miniclusters between $[10^{-16},\,10^{-12}]~M_\odot$, and $\gamma = -0.5$ for merged minihalos up to $5\times 10^{-7}~M_\odot$. We discuss N-body results and justify these inputs in the supplemental material (S.M.)~Sec.~I, which includes Refs.~\cite{1958ApJ...127...17S,Springel:2020plp,Jiang:2014nsa,vandenBosch:2004zs,Dehnen:1996fa,Bissantz:2001wx,Binney:1996sv,Zagorac:2022xic,Nori:2020jzx,Chan:2021bja,Schive2014_PRL,Du:2023jxh,Chen:2020cef,Levkov2018,Visinelli:2017ooc,Navarro:1995iw,Dai:2019lud,Dandoy:2022prp}. 

\noindent \textbf{\textit{Monte Carlo simulation}}.---After drawing a sample of miniclusters, we propagate their orbits around the galaxy. We are interested here in the set of all possible galactic orbits that end in the Solar neighbourhood today. Velocities are sampled according to the Standard Halo Model (SHM) by drawing from an isotropic 3D Gaussian with width $\sigma_{\rm halo} = v_{\rm circ}/\sqrt{2}$ where $v_{\rm circ} = 233$ km/s is the circular speed of the Solar orbit~\cite{Evans:2018bqy}. We truncate the velocity distribution at $v_{\rm esc} = 528$~km/s~\cite{Piffl:2014mfa} to discount orbits unbound to the MW. We integrate orbits over a duration $t_{\rm MW}=13.5$~Gyr using \texttt{galpy}~\cite{Bovy_2015}, adopting the commonly-used potential ``MWPotential2014''.

For each orbit, we evaluate the variation in the local stellar number density $n_{\star}(t)$ felt by the minicluster. Our galactic model includes a central bulge and the thin/thick disks---described in S.M.~II. The total number of stars encountered by the minicluster is the integral,
\begin{equation}
    N_{\rm enc}=\int_0^{t_{\rm MW}}{\rm d}t~n_{\star}(t)v(t)\pi b^2_{\rm max}\, ,\label{eq:nenc}
\end{equation} 
where $v(t)$ is the minicluster velocity, and $b_{\rm max}$ some maximum impact parameter between the minicluster and a star. We want to set $b_{\rm max}$ to be as large as possible to capture all possible disruption but without being so large as to add unnecessary computational burden by having large numbers of negligible encounters. As in Ref.~\cite{Kavanagh:2020gcy}, we find that $b_{\rm max} = 0.1$~pc strikes a good balance. Increasing this number by a factor of 10 does not change our results. The $N_{\rm enc}$ encounters are labelled by a set of encounter times $\{ t^i_{\rm enc}\}$, drawn from a probability distribution proportional to the integrand in Eq.(\ref{eq:nenc}). The impact parameter, $b$ of each encounter, are drawn randomly from inside a circle of radius $b_{\rm max}$.

Stars appear to miniclusters as point-like objects, which means we may work in the \emph{distant tide} approximation, where the impact parameter between the minicluster and a star is much larger than the minicluster radius, $b\gg R$. The energy injected by an encounter with a star of mass $M_\star$ under this approximation is~\cite{Green:2006hh,Schneider:2010jr},
\begin{equation}
         \Delta E\simeq \left(\frac{2GM_\star}{b^2 v_{\rm rel}}\right)^2\frac{M\langle R^2\rangle}{3}\,,\label{eq:distide}
\end{equation} 
where $v_{\rm rel}$ is the minicluster-star relative velocity, and $\langle R^2\rangle$ is the minicluster mean-squared radius (see S.M.~III for how the latter is calculated). 
Most encounters inject $\Delta E\ll E_b$ so we must deal with \emph{perturbations}, which do not totally disrupt the minicluster in one go, but lead to a series of mass losses $\Delta M^i$. The procedure then is to execute $i = 1,...,N_{\rm enc}$ successive perturbations over the minicluster's orbit, repeatedly updating the mass and radius that it relaxes to, until it is either fully unbound or we reach the end of the orbit. The recipe for this procedure is given in S.M.~III. For computational efficiency, we group together the very large number of encounters occurring during a disk-crossing event, and only allow the minicluster to relax to a new profile and radius over its relaxation time between disk-crossings. This is justified here because the relaxation time [\mbox{$t_{\rm rel} \sim (G \rho_{\rm mc})^{-1/2}\sim \mathcal{O}(10\, {\rm Myr})$}] is shorter than the timescale between disk-crossings where major encounters occur: $\mathcal{O}(10-100\, {\rm Myr})$.

We find that isolated miniclusters are relatively stable against disruption, and the majority survive with some mass intact. The high-mass merged miniclusters are much more easily disrupted, with around half of them losing $>99\%$ of their mass by today.

\noindent \textbf{\textit{Stream formation}}.---When a certain amount of the minicluster's mass is stripped, where does it go? Although this unbound mass shell has effectively evaporated off of the minicluster, it will still retain its host's centre-of-mass orbital velocity, and so continue along the same orbit in the vicinity. The tidal field of the Milky Way will continue to act on these unbound particles, causing the debris to elongate into a stream in the direction of the orbit. This is seen generically for tidally disrupted structures at all scales in astrophysics, but has also been simulated for DM microhalos on the scales we are interested in here in Ref.~\cite{Schneider:2010jr}. We use this study as inspiration for our stream model. We assume that an unbound mass shell turns into a tidal stream, with a leading tail that advances beyond the original host minicluster and a trailing tail that lags behind. By today, a given mass shell evaporating in some encounter at time $t^i_{\rm enc}$ will have turned into a stream of \textit{minimum} length given by the minicluster velocity dispersion~\cite{Schneider:2010jr}, $\ell^i_{\rm str} \gtrsim \sigma^i_{\rm mc} (t_{\rm MW}-t^i_{\rm enc})$.
It is important to note that the stream will likely be longer than this today due to the energy injection and further tidal heating, potentially by a factor of 10 compared to $\sim \sigma_{\rm mc} t$ ~\cite{Schneider:2010jr}. We explore levels of additional elongation in S.M~IV, but for reasons that will become clear, taking the smallest possible length that the streams could grow to is the most conservative option for the question we are trying to answer. 

To model the stream formation, we will make the assumption that the mass lost in each tidal disruption event goes into a cylinder the same radius as the original minicluster, $R$, and with a Gaussian density profile running along the stream. The total stream is then comprised of the sum of all of these individual evaporated mass shells occurring at different times. We can parameterise this in terms of a function $\rho_{\rm str}(\ell)$,
\begin{equation}
    \rho_{\rm str}(\ell) = \sum_{i=1}^{N_{\rm enc}} \frac{\Delta M^i}{\pi R_{\rm mc}^2 \sqrt{2\pi (\ell^i_{\rm str})^2}} \exp\left( -\frac{\ell^2}{2(\ell^i_{\rm str})^2}\right) \, ,\label{eq:rhoell}
\end{equation} 
where $\ell$ is a coordinate that runs along the stream. The length scales for the miniclusters (weighted by the $\Delta M$ in each segment) are in the range $7.3^{+3.2}_{-4.0}$~pc for merged miniclusters and $0.28^{+0.34}_{-0.2}$~pc for isolated miniclusters. 

\noindent\textit{\textbf{The local density in minicluster streams}}.---We will now present an argument for the degree to which the local DM density at our position in the galaxy is replenished by the disruption of the miniclusters into long streams. First, imagine a sphere around the Sun of radius 100~pc. This is the order-of-magnitude scale within which we have strong evidence for a galactic DM density of $\rho_{\rm DM}\approx 0.4\,$GeV/cm$^3\approx 0.01\,M_\odot\,{\rm pc}^{-3}$~\cite{deSalas:2020hbh}. Within this sphere, the total mass of DM is $M_{\mathrm{DM}}=4.2\times 10^{4} M_{\odot}$. From Ref.~\cite{Eggemeier:2022hqa}, we know around $f_{\rm void} = 8\%$ of this should be comprised of an ambient density of unbound axions---the minivoids. Therefore $M_{\rm DM}(1-f_{\rm void})$ is the total mass of axions that were initially bound in miniclusters inside this volume. Using our baseline mass function, this implies a total of around $N_{\rm mc} \sim 10^{14}$ isolated+merged miniclusters.

Let us assume the final volume occupied by each stream after the disruption process is $V_{\rm str} \approx \pi R_{\rm mc}^2 l_{95}$. We define $l_{95}$ to be the length within which 95\% of the stream mass is contained, calculated using Eq.(\ref{eq:rhoell}). If we have $N_{\rm mc}$ miniclusters inside a volume $V = 4/3 \pi r_{\rm local}^3$ and those miniclusters each occupy a volume $V_{\rm str}$ after disruption, the expected number of streams overlapping a random point inside the sphere is then,
\begin{equation}
    \langle N_{\rm str} \rangle = \sum_{j = 1}^{N_{\rm mc}} \frac{V^j_{\rm str}}{V_{\rm local}}\, , \label{eq:sum}
\end{equation}
where the ratio of the two volumes gives us the probability that a particular stream overlaps our chosen point. For our baseline set of assumptions, we find that this number is $246\pm15$ (statistical error). Varying our assumptions---for example the density profiles, the concentration of the NFW merged miniclusters, and/or accounting for additional stream heating---we obtain numbers in the range $\sim 200$ to $6000$, see S.M.~IV. 

With this number, we can now re-sample from our distribution of streams $\mathcal{O}(10^2-10^3)$ times, with probabilities weighted by $V_{\rm str}^i$. For each stream, we also randomly choose our position within it $\ell \in [-l_{\rm 95}/2,l_{\rm 95}/2]$ to get the value of the DM density that it contributes from Eq.~\eqref{eq:rhoell}. Repeating this process many times, we find that the sum of all the individual densities adds up to a total
\begin{equation}\label{eq:rhostr}
    \frac{1}{\rho_{\rm DM}} \sum_{i=1}^{N_{\rm str}} \rho^i_{\rm str}(\ell^i_\odot) = 0.81 \pm 0.06  \, ,
\end{equation}
where \mbox{$\rho_{\rm DM}\approx 0.4\,$GeV/cm$^3$} is the usual DM density inferred on much larger scales. So when added to the 8\% of the DM density originally filling the minivoids, this is a replenishment of almost 90\% of $\rho_{\rm DM}$---a substantial improvement in the prospects for direct detection.

We emphasise here that this final number is generally insensitive to many of our cruder assumptions. The first point to state is that the debris from disrupted miniclusters is dominated almost entirely by the \textit{merged} ones. These streams constitute the most DM by mass, and have the highest probability of intersecting our position because they cover more volume.

The second key point is that the expected value of the DM density (Eq.~\ref{eq:rhostr}) is robust against re-parameterisations as long as we are in the regime when $N_{\rm str}\gg 1$. That said, the number of expected streams is somewhat arbitrary, but given that it is very safely $\gg 1$ for any reasonable choice of parameters, we show explicitly in S.M.~IV that the typical (i.e.~median, or mean) DM density they add up to does not depend on this number because we have conserved the total DM mass. Instead, the number of streams affects the \textit{variance} in $\rho_{\rm str}$. It can also be shown (see S.M.~IV) that as long as the minicluster mass and radius are related like $R\propto M^{1/3}$ then the mass-dependence drops out entirely, meaning assumptions about the mass function also do not affect $N_{\rm str}$, leaving only a dependence on the NFW concentration parameter. 

Quantities that affect $N_{\rm str}$ include the typical radii that the miniclusters are truncated at (related to the concentration parameter for NFW profiles), and the extent to which streams are additionally elongated beyond the minimal expectation $\sigma_{\rm mc}t$. If we allow the miniclusters to be larger in radius, or the streams to be longer, then the typical number of streams overlapping at our position increases to $\mathcal{O}(10^3)$---however this means that the variance in the mean density our position ends up \textit{smaller}. Because of this, our baseline parameterisation is the most conservative---allowing for processes that can further strip the miniclusters or elongate streams only acts to suppress the variation in the value stated in Eq.(\ref{eq:rhostr}). Figure S3 in the S.M. shows the impact of these uncertainties quantitatively.

Put together, the only uncertainty remaining that could change our results substantially is if the merged miniclusters do not continue to evolve towards smooth NFW profiles: If instead some of them remain as ``clusters of miniclusters'', then it is possible that they are more resilient. We discuss this issue in S.M.~V. The takeaway from our N-body simulation halo finder is that the majority of the axions inside merged miniclusters are indeed attached to their host halo as opposed to internal subhalos. 

\begin{figure}
    \centering
    \includegraphics[width=0.45\textwidth]{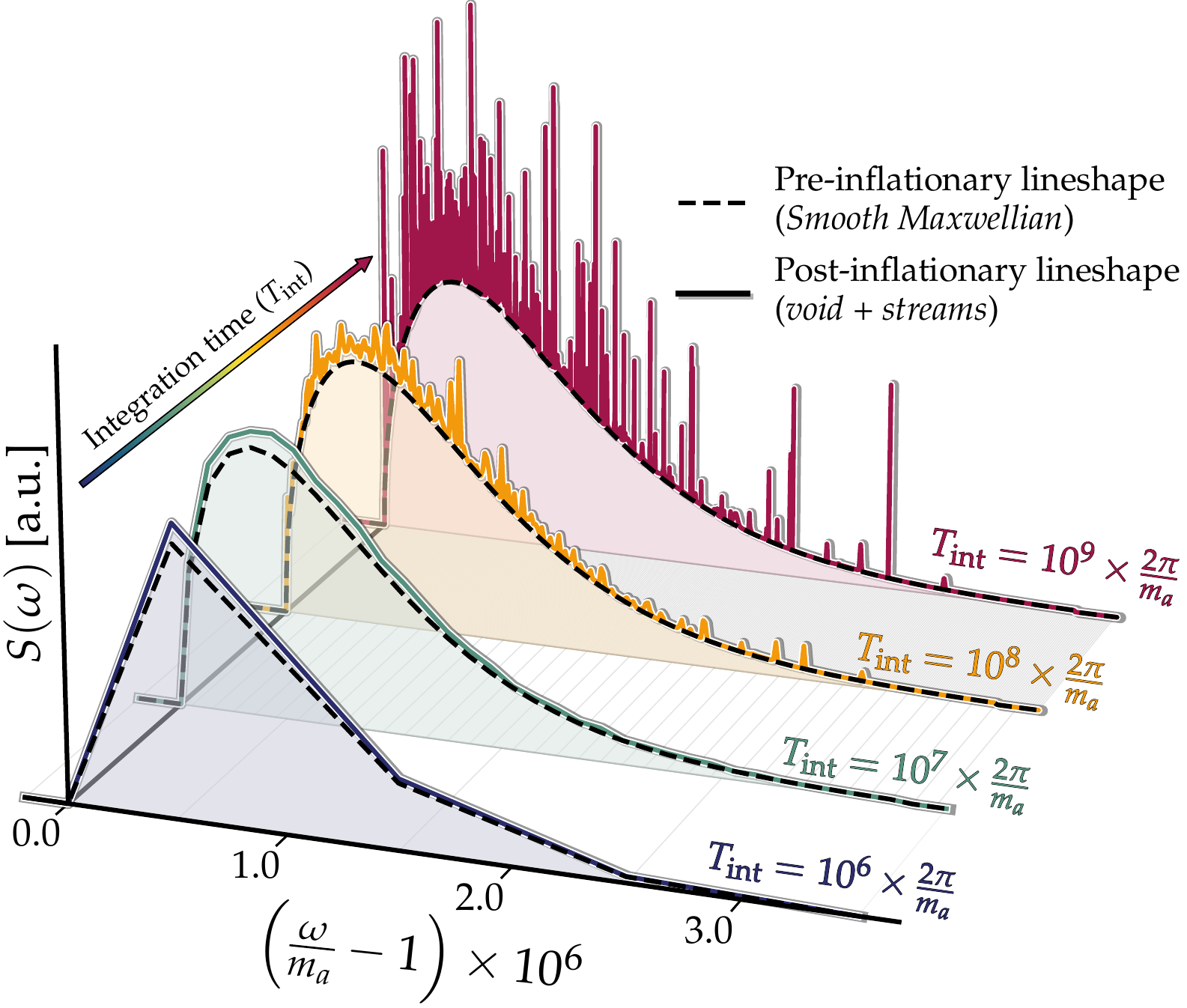}
    \caption{An illustration of how minicluster streams would manifest in the experimental signature of axions called the ``lineshape''. The frequency resolution is inversely proportional to the duration of the observation, $T_{\rm int}$, from which this spectrum is obtained via a discrete Fourier transform. If $T_{\rm int}$ is short, then the lineshape is indistinguishable from a smooth halo. However, samples longer than $\sim 10^8$ oscillation periods have sufficient resolution to identify the streams.}
    \label{fig:signal}
\end{figure}

\noindent\textit{\textbf{Impact on axion haloscopes}}.---We now know how many streams we expect to have around us at any one time, and so we can draw a sample of this size and predict how this would look in a haloscope. 
First, we build the velocity distribution of axions. In the voids, we can safely assume the halo is described by the smooth, fully-virialised SHM~\cite{Evans:2018bqy} modelled as an isotropic Gaussian with width \mbox{$\sigma_{\rm void} = v_{\rm circ}/\sqrt{2} \approx 165$ km/s}. An experiment observes this after a Galilean boost into our frame of reference, moving at a velocity $\mathbf{v}_{\rm lab}(t)$ with respect to the Galactic centre. We take \mbox{$\mathbf{v}_{\rm lab} = (11.1,235.2,7.3) \,{\rm km/s} + \mathbf{v}_{\oplus}(t)$}~\cite{Evans:2018bqy} in the same galactocentric coordinate system as Fig.~\ref{fig:main}, where $\mathbf{v}_{\oplus}(t)$ is the Earth velocity, see e.g.~\cite{McCabe:2013kea,Mayet:2016zxu}.

Following past literature~\cite{Freese:2003tt, OHare:2014nxd, Foster:2017hbq, OHare:2018trr, OHare:2019qxc, Ko:2023gem}, we take the stream velocity distribution to have the same Gaussian form, except we boost by $\mathbf{v}_{\rm lab} - \mathbf{v}_{\rm str}$ to account for the stream's velocity, and set the width to $\sigma_{\rm str}$. Because the values of $\sigma_{\rm str}$ are small, the stream signals are extremely narrowband in frequency.

We now relate this velocity distribution to the signal measured by a typical axion haloscope known as the \textit{lineshape}. The axion field's oscillations are highly coherent, in keeping with its description as cold DM. Within a ``coherence time'', the axion field will appear to oscillate at a single frequency $\omega \approx m_a(1+v^2/2)$ where $v\sim10^{-3}c$ is some speed drawn from $f(v)$. 
This frequency will then evolve on timescales longer than coherence time, depending on the spread in velocities: $\tau_{\rm coh} \sim 1/m_a \sigma^2_{\rm void} \sim 10^{6} \,m^{-1}_a \sim {\rm 0.01 \,ms}\,(100\mu{\rm eV}/m_a)$. If oscillations are measured over timescales much longer than this, then the discrete Fourier transform of that measurement will have a spectrum related to the distribution of component frequencies. For a measurement time, $T_{\rm int}$, the power spectrum $S(\omega)$ has a frequency resolution of $\Delta \omega = 2\pi/T_{\rm int}$. Therefore, if $T_{\rm int}> 10^6 \times 2\pi/m_a$ (i.e.~longer than a million oscillation periods) then we expect the axion's lineshape to be resolved~\cite{Turner:1990qx,Derevianko:2016vpm,Foster:2017hbq,OHare:2017yze}.

To illustrate this, we construct a simplified version of a signal power spectrum by discretising the distribution of frequencies into bins of width $\Delta \omega$~\cite{Foster:2017hbq, OHare:2017yze, Knirck:2018knd}. 
Figure~\ref{fig:signal} compares the smooth Maxwellian case (as in, for example, pre-inflationary axions) against the case where there are streams in the signal. The two will become strikingly different once the signal is integrated for timescales longer than $10^8$ oscillation periods.

As well as showing up as sharp features in short-duration measurements, the lineshape will also evolve in amplitude over human timescales as we traverse the streams. Sampling over all of our streams and randomising our trajectory through them, we find that they would typically persist for around $\Delta t= 30^{+ 23}_{- 11} \, {\rm years}$ (median and 68\% containment). However, since we expect $10^2$--$10^3$ present in the lineshape at one time, the timescale over which the signal is expected to vary is on the order of weeks. Ultra-narrowband axion signals like these are already being searched for by haloscope collaborations~\cite{Duffy:2005ab, ADMX:2006kgb, ADMX:2023ctd}, so in light of our results, we advocate that these efforts continue.

\noindent{\textbf{\textit{Conclusions}}}.---We have evaluated the extent to which axion haloscopes are doomed to never discover the axion in the post-inflationary scenario because axions find themselves bound up inside of small substructures. By modelling the tidal disruption of these miniclusters by stars, we have found that the density of DM around us in the solar neighbourhood is refilled to around 80\% of the 100-pc-scale average density value $\rho_{\rm DM} \approx 0.4$~GeV~cm$^{-3}$ usually adopted by experiments. Combined with an estimate of the leftover density of DM in minivoids~\cite{Eggemeier:2022hqa}, this boosts the signal up to an impressive 90\% of the commonly assumed value. In other words: axion haloscopes may not be doomed. 


\noindent{{\textit{Acknowledgements}}}.---
We give special thanks to Chris Gordon and Ian DSouza for pointing out critical typos in the original version of this paper. We also thank Benedikt Eggemeier, David Marsh and Yvonne Wong for helpful discussions. CAJO is funded via the Australian Research Council, grant number DE220100225.
The work of J.R. is supported by Grants PGC2022-126078NB-C21 funded by MCIN/AEI/ 10.13039/501100011033 and “ERDF A way of making Europe” and Grant  DGA-FSE grant 2020-E21-17R Aragon Government and the European
Union - NextGenerationEU Recovery and Resilience Program on `Astrofísica y Física de Altas Energías' CEFCA-CAPA-ITAINNOVA. This article is based upon work from COST Action
COSMIC WISPers CA21106, supported by COST (European Cooperation in Science and Technology). 
We acknowledge the work of G.P.'s PhD thesis submitted in September
2023.

\bibliography{axions.bib}
\bibliographystyle{bibi}

\clearpage
\newpage
\maketitle
\onecolumngrid
\begin{center}
\textbf{\large Axion minicluster streams in the solar neighbourhood} \\ 
\vspace{0.1in}
{ \it \large Supplemental Material}\\ 
\vspace{0.05in}
{Ciaran A.~J.~O'Hare \ Giovanni Pierobon, and \ Javier Redondo}
\end{center}
\onecolumngrid
\setcounter{equation}{0}
\setcounter{figure}{0}
\setcounter{table}{0}
\setcounter{section}{0}
\setcounter{page}{1}
\makeatletter
\renewcommand{\theequation}{S\arabic{equation}}
\renewcommand{\thefigure}{S\arabic{figure}}

\section{Minicluster properties from early-Universe simulations}\label{app:sims}

In this section, we give further details on the simulations that inspire our initial population of miniclusters and detail the structural parameters that are necessary to model their subsequent disruption. As described in Refs.~\cite{Eggemeier:2022hqa,Pierobon:2023ozb} we evolve realistic configurations of the axion DM field within the redshifts $10^{16}\lesssim z\lesssim 10^2$. 

The distribution of miniclusters can either be predicted using the spectrum of density fluctuations $\Delta^2$ left behind following the collapse of the axion string-wall network, or they can obtained directly by performing an N-body simulation. Recently, Ref.~\cite{Pierobon:2023ozb} demonstrated how both the mass distribution and internal structures of miniclusters vary when adopting different treatments of the early Universe dynamics---in particular the modelling of axion strings. Due to these outstanding uncertainties, current simulations~\cite{Eggemeier:2019jsu,Xiao:2021nkb,Pierobon:2023ozb} are unfortunately unable to conclusively identify a universal density profile for all forming miniclusters. For this study, we have chosen to adopt the early Universe modelling of direct simulations at large string tension (dubbed high-$\kappa$) from Ref.~\cite{Pierobon:2023ozb}, which best describes the string-wall network with respect to previous works.

\begin{figure}
    \centering
    \includegraphics[width=0.75\textwidth]{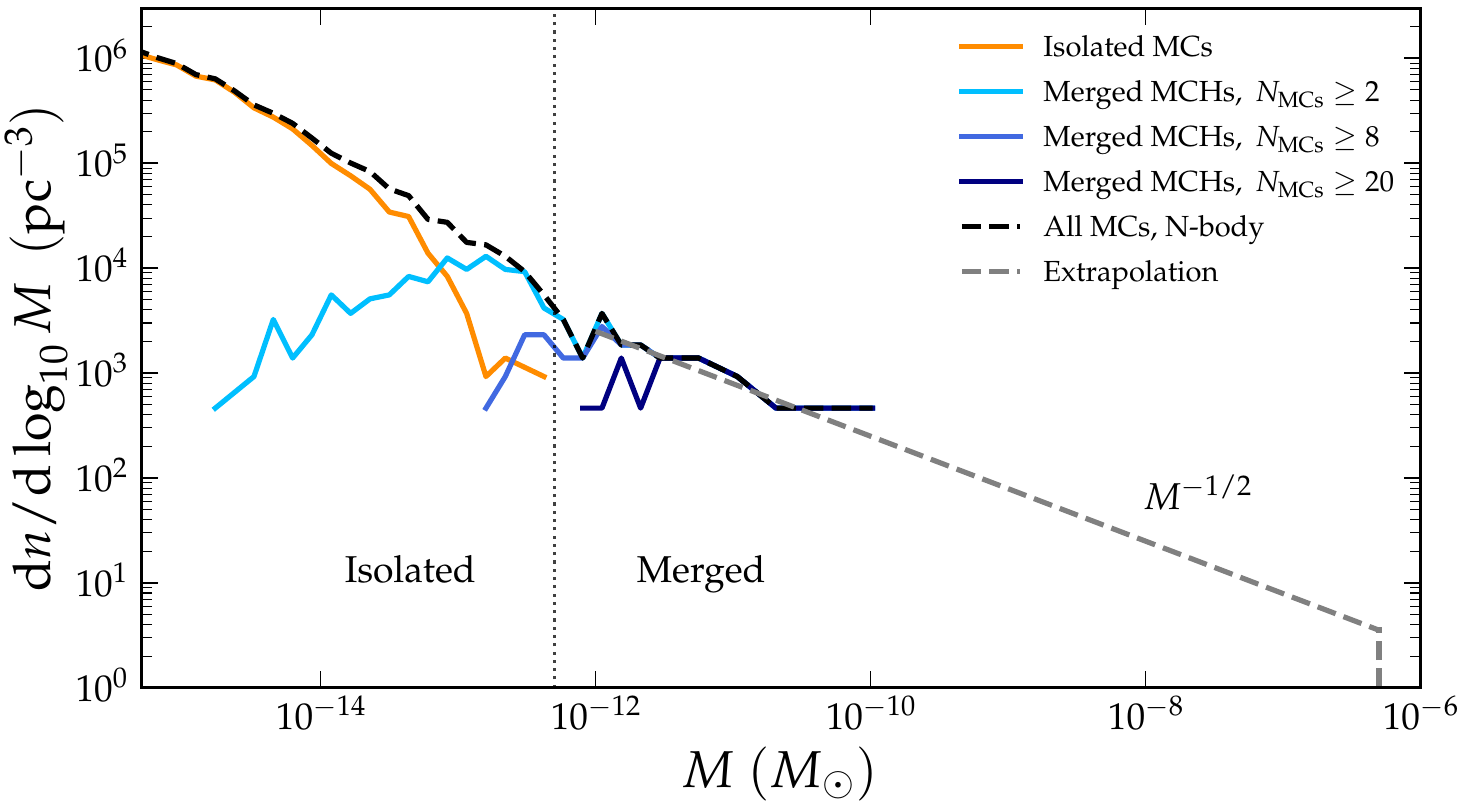}
    \caption{Halo mass functions resulting from N-body simulations. The isolated minicluster mass spectrum is derived from simulations that had initial conditions calculated by a high-string-tension lattice simulation (high-$\kappa$, as in Ref.~\cite{Pierobon:2023ozb}). On the other hand, the high-mass extrapolated mass function for hierarchically merged miniclusters is inspired by N-body simulations evolving miniclusters up to $z=19$. This latter regime, which has an initial power spectrum well-modelled as Gaussian white noise, extends up to $\sim 5 \times 10^{-7}~M_\odot$. The total mass function for all miniclusters and minicluster halos (MCHs) is shown as a dashed line, while the coloured lines show the individual mass functions for minicluster halos comprised of different numbers of individual miniclusters. Our final results are heavily dominated by the upper end of this mass function comprised of the mergers of many miniclusters, however the dependence on details like the maximum mass and the exact power law turn out to not have a significant impact on our results.}
    \label{fig:hmf}
\end{figure}

Uncertainties aside, in more general terms however, recent N-body simulations are able to show that the NFW profile fits well the high-mass end of the mass function corresponding to hierarchical mergers of smaller miniclusters. Whereas on the low-mass end, consisting of halos that remain isolated after their initial collapse, power-law profiles fit much better. These two types of minicluster can be seen in Fig.~\ref{fig:hmf} where we display the halo mass function taken from the N-body simulations presented in Ref.~\cite{Pierobon:2023ozb}. The non-linear regime refers to those miniclusters that promptly collapse and then do not undergo significant mergers, whereas the high mass tail, which grows out of the hierarchical merger of many initially isolated miniclusters, has a spectrum well-described by Gaussian white noise, hence why it is referred to the white-noise regime. We describe how we model each of these now in turn. The high-mass extrapolation is inspired by the N-body simulations, which follow the miniclusters to even higher redshifts of $z=19$, when they fall into the first galactic halos.

\subsection{Isolated miniclusters}
Isolated miniclusters are defined to be the early-forming ($z\gtrsim z_{\rm eq}$) bound objects associated with the large nonlinear overdensities left by inhomogeneities in the axion field, which in turn result directly from the collapse of the string-wall network. We find that isolated MCs occupy the mass distribution for masses $M\lesssim 10^{-12}~M_{\odot}$ and have radii $R\lesssim 10^{-6}$ pc.

Isolated miniclusters can be fit well using a power-law profile in terms of an index $\tilde{\alpha}$,
\begin{equation}
    \rho_{\rm PL}(r)=Cr^{-\tilde{\alpha}},\quad \quad C=\frac{3-\tilde{\alpha}}{3}\langle\rho\rangle R^{\tilde{\alpha}}
, 
\end{equation}

where $\langle\rho\rangle$ is the average energy density of the minicluster. In our simulations, they typically exhibit power-law profiles with index $\tilde{\alpha}=9/4$ at $z\sim z_{\rm eq}$. However at $z\sim 500$ the value shifts to an average $\langle\tilde{\alpha}\rangle=2.71$. Densities are sampled from a Gaussian distribution built using the mean and standard deviation of simulation data of small-mass miniclusters. We assume the isolated miniclusters occupy the range $M\in[10^{-16},10^{-12}]~M_{\odot}$, and have a mass function (number density of miniclusters versus mass) given by $\textrm{d}n/\textrm{d}\log M\propto M^\gamma$, with $\gamma = -0.8$. 

Such halos represent about 70\% of the total by \textit{number}---resulting as a consequence of the mass cutoff that divides what constitutes an isolated versus a merged minicluster. This is seen in high-$\kappa$ N-body simulations, first described in Ref.~\cite{Pierobon:2023ozb}. We mention these details for clarity, but as discussed in the main text, and above, we ultimately find that our key result is not impacted by our modelling of the isolated miniclusters, and our results are unchanged even if we make extreme departures from any of the assumptions we have made about them.

\subsection{Merged miniclusters}
Much more important are the merged minicluster halos, forming in the white-noise regime, corresponding to fluctuations on scales $k\ll L_1^{-1}$. Defining $M_1$ as the characteristic mass defined as \footnote{We define $M_1$ as in Ref.~\cite{Dai:2019lud}.
}
\begin{align}
        M_1&=\frac{4\pi}{3}\rho_a(\pi L_1)^3\\
        &\simeq 1.76\times 10^{-10}\,M_\odot~\left(\frac{100~\mu{\rm eV}}{m_a}\right)^{0.51}.
   \end{align}
where $m_a$ is the axion mass, merged miniclusters will have masses $M\gtrsim M_1$, for which the spectrum is well-described by Gaussian white-noise~\cite{Enander:2017ogx}. 

Merged minicluster halos around and above the characteristic mass evolve towards a Navarro-Frank-White (NFW)~\cite{Navarro:1995iw} profile, which is given by
\begin{equation}
    \rho_{\rm NFW}(r)=\frac{\rho_s}{(r/r_s)(1+r/r_s)^2}, 
\end{equation} 
with $r_s,\rho_s$ scale radius and scale density, respectively. To model this population, we use numerical results of the white-noise N-body simulations from Ref.~\cite{Xiao:2021nkb}, which found a scale radius and concentration as a function of the minicluster mass and redshift, $z$.
\begin{equation}
    r_s(M)=3.7 h^{-1} \, {\rm mpc} \left(\frac{A_{\Delta^2}M_1}{10^{-11}~M_{\odot}h^{-1}}\right)^{-1/2}\left(\frac{M}{10^{-6}~M_{\odot}h^{-1}}\right)^{5/6}, \label{eq:9scaler}
\end{equation}  
\begin{equation}
    c(z)=\frac{R}{r_s}=\frac{1.4\times 10^4}{(1+z)}\left(\frac{M}{A_{\Delta^2}M_1}\right)^{-1/2}.\label{eq:9conc}
\end{equation}
From this, the radius at which the profile is truncated to contain the mass $M$ can be found to be~\cite{Xiao:2021nkb}
\begin{equation}
    R(M) \equiv  R_{\rm 10}\left(\frac{M}{10^{-10}~M_{\odot}}\right)^{1/3} \simeq 0.55\,{\rm mpc}~\left(\frac{M}{10^{-10}~M_{\odot}}\right)^{1/3}.\label{eq:mergedR}
\end{equation} 
which we adopt to model our minicluster, taking $z=19$. Note that the overall scale for this radius is almost a factor of three larger than that assumed by Kavanagh et al.~\cite{Kavanagh:2020gcy} in their minicluster disruption study. Their value is equivalent to a concentration parameter of only $c = 100$, which they argue would be a result of rapid initial truncation by tidal stripping as the miniclusters originally fell into the MW halo--this procedure we will also be implementing in our mass-loss pipeline. Nonetheless, this overall truncation scale of the profile ends up impacting some of our results. Instead of parameterising this uncertainty in terms of $c$, which does not directly enter our results, we instead parameterise the truncation \textit{radius} by varying the prefactor in the expression above. We define this to be $R_{10}$ and will show results as a function of this quantity in Sec.~\ref{app:uncertainties}.

For our NFW halos, the individual mean densities are sampled from a normal distribution with mean and variance according to the N-body data~\cite{Pierobon:2023ozb}. We sample masses in the range \mbox{$M\in[10^{-12},5\times 10^{-7}]~M_{\odot}$} according to halo-mass function ${\rm d}n/{\rm d}\log M \propto M^{\gamma}$ with $\gamma=-0.5$. The upper mass cut-off has been estimated as $\sim 0.03 \langle \rho\rangle L_1^3 z_{\rm eq}^2$ from the Press-Schechter estimate of the variance obtained from numerical simulations in \cite{Pierobon:2023ozb}. 

Before moving on, it is worth remarking here briefly that our results are largely insensitive to the question of whether or not miniclusters host small solitonic cores called axion stars~\cite{Visinelli:2017ooc}. Numerous recent studies have suggested that all gravitationally bound structures of a bosonic field ought to develop these cores~\cite{Levkov2018, Eggemeier:2019jsu, Chen:2020cef}, which may survive and even grow during mergers~\cite{Du:2023jxh}. Assuming the core-halo mass relation seen in simulations of the Schrodinger-Poisson system~\cite{Schive2014_PRL} is universal (i.e.~down to minicluster masses~\cite{Eggemeier:2019jsu}), we can evaluate it at the present day to yield the following relationship with the minicluster mass,
\begin{equation}
    M_{\rm star}=5.58\times 10^{-17}\, M_\odot\left( \frac{100 \, \mu {\rm eV}}{m_a} \right) \left( \frac{M}{10^{-10}\,M_\odot} \right)^{1/3}
\end{equation}
Some intrinsic diversity away from this relation has been seen in some simulations, but we are primarily interested in the ratio between the core and halo mass at an order-of-magnitude level~\cite{Nori:2020jzx,Chan:2021bja,Zagorac:2022xic}. 

Axion stars have presented a potentially problematic theoretical uncertainty in previous studies that focused on the survival of miniclusters. Since the axion star could potentially prevent miniclusters from being totally stripped. However, we notice here that the total mass contained in the axion star is vastly subdominant, and all of our results depend dominantly on the loosely bound outer layers of the high-mass merged miniclusters which are stripped early on in the evolution. Hence we do not need to worry about whether or not miniclusters host axion stars, and accounting for presence in the cores of our miniclusters, or implementing cuts on stripped miniclusters which would host unphysically large axion stars (as in Ref.~\cite{Kavanagh:2020gcy}), would not change any of our results.

\section{Galactic stellar density model}\label{app:stardens}
The number of stars encountered within an impact parameter of $b_{\rm max}$ is given by the integral along the minicluster trajectory:
\begin{equation}
    N_{\rm enc}=\int_0^{t_{\rm MW}}{\rm d}t~n_{\star}(t)v(t)\pi b^2_{\rm max}\, .
\end{equation} 
To get the stellar number density, $n_{\star}(t)$ we write down a model for the galactic stellar mass density and choose a characteristic star mass $M_{\star} = 1~M_\odot$. Because the disruption probability is proportional to the number density of stars times their mass, including a detailed stellar mass function is not expected to change our conclusions if we keep the volume and total mass fixed~\cite{Dokuchaev}--and in any case, $1~M_\odot$ is a very typical star mass for standard stellar initial mass functions.

Our stellar density model, as in Refs.~\cite{Kavanagh:2020gcy,Dokuchaev}, consists of a bulge and disk, of which the latter has two components, labelled as thick and thin:
\begin{equation}
    \rho_{\star}=\rho_{\rm bulge}+\rho^{(t)}_{\rm disk}+\rho^{(T)}_{\rm disk}\,.\label{eq:sstarden}
\end{equation} 
The bulge density profile is modelled as a truncated power-law~\cite{Binney:1996sv,Bissantz:2001wx,Kavanagh:2020gcy} 
\begin{equation}
     \rho_{\rm bulge}(r,z)=\rho_{{\rm bulge},0}\frac{\exp(-r'/r_{\rm cut})^2}{1+(r'/r_0)^{1.8}},\quad r'=\sqrt{(r^2+4z^2)}
\end{equation}
where $r = \sqrt{x^2 + y^2}$ is the cylindrical radius in the disk-plane, and $z$ the height above the disk plane. The central density is $\rho_{{\rm bulge},0}\simeq 99.5~M_{\odot}/{\rm pc}^3$, and the radial scales are $r_0=0.075$ kpc and $r_{\rm cut}=2.1$ kpc.

Both the thick and the thin stellar disk profiles are modelled by a double exponential~\cite{Dehnen:1996fa} 
\begin{equation}
    \rho_{\rm disk}^{t,T}(r,z)=\frac{\Sigma^{t,T}_\star}{z^{t,T}}\exp\left(-\frac{r}{r^{t,T}}-\frac{Z}{z^{t,T}}\right),
\end{equation}
where $r^{t,T},z^{t,T}$ are the scale radius and height, and $\Sigma^{t,T}_\star$ are the stellar surface densities:
\begin{align}
    z^{t}&=0.3~{\rm kpc},\quad r^{t}=2.90~{\rm kpc}\quad \Sigma^{t}_\star=816.6~M_{\odot}~{\rm pc}^{-2},\\
    z^{T}&=0.9~{\rm kpc},\quad r^{T}=3.31~{\rm kpc}\quad \Sigma^{T}_\star=209.5~M_{\odot}~{\rm pc}^{-2}.
\end{align} 
Our results are not sensitive to the detailed parameterisation of the stellar density, and varying any of these numbers within up to a few tens of percent leaves our results generally unaffected.

\section{Minicluster disruption computation}\label{app:disruption}
Here we provide a brief account of the disruption computation, though for further discussion of the motivation and justification behind this treatment, we refer to Ref.~\cite{Kavanagh:2020gcy}.

Once an encounter time ($t_{\rm enc}$), star mass ($M_\star$), and impact parameter ($b$) have been chosen, the first step is to compute the energy injected into the minicluster by that encounter. Under the distant tide and the impulse approximations assumptions, the energy injection can be written as~\cite{1958ApJ...127...17S}:
\begin{equation}
         \Delta E\simeq \left(\frac{2GM_*}{b^2 v_{\rm rel}}\right)^2\frac{M\langle R^2\rangle}{3},
\end{equation} 
where $\langle R^2\rangle$ is the minicluster mean-squared radius. This can be expressed in terms of a structural parameter, $\alpha$,
\begin{equation}
    \alpha^2=\frac{\langle R^2\rangle}{R^2}=\frac{4\pi}{MR^2}\int_0^{R}{\rm d}rr^4\rho(r)\,.\label{eq:alp2} 
\end{equation} 
For an NFW minicluster with concentration $c=100$, Refs.~\cite{Kavanagh:2020gcy,Shen:2022ltx} find $\alpha^2=0.13$. For the power-law profile, we use $\alpha^2 = 0.27$~\cite{Kavanagh:2020gcy}.

The energy injection must then be compared to the binding energy of the minicluster, $E_b$, which is given by the familiar $\sim GM^2/R$ up to some factor $\beta$, that encodes details of the internal structure. We define it via,
\begin{equation}
    E_b\simeq \beta\frac{GM^2}{R},  \label{eq:binnd}
\end{equation}
where the structural parameter is (see, e.g., Refs.~\cite{Kavanagh:2020gcy,Dandoy:2022prp}.) 
\begin{equation}
    \beta=\frac{4 \pi R}{M{ }^2} \int_0^{R} \mathrm{d} r r \rho(r) M_{\mathrm{enc}}(r)
\end{equation}
where $M_{\rm enc}(r)$ is the mass enclosed inside radius $r$. If the energy injected into the minicluster by the star's tidal field exceeds the minicluster's binding energy, we expect the minicluster to lose and order-1 fraction of its mass~\cite{Shen:2022ltx}. However, since most encounters inject $\Delta E\ll E_b$, we deal with \emph{perturbations} giving rise to small mass losses $\Delta M$. 

In the distant tide approximation, the normalised injected energy increases proportionally to $R^2$: 
\begin{equation}
    \Delta\mathcal{E}(R)=\frac{\Delta E}{M}\frac{R^2}{\langle R^2\rangle},
\end{equation} 
where $\mathcal{E} = -E/m_a$ is the energy per unit mass. The mass that is then left unbound to the cluster following this injection is then given by the total mass in axions with energies smaller than $\Delta\mathcal{E}$~\cite{Kavanagh:2020gcy}
\begin{equation}\label{eq:massloss}
    \Delta M(\Delta E) \equiv M(<\Delta \mathcal{E})=\int_{\mathcal{E}<\Delta \mathcal{E}(R)} \mathrm{d}^3 \mathbf{x} \, \mathrm{d}^3 \mathbf{v} f(\mathcal{E})
\end{equation}
In our computation, we use a numerical interpolation of this mass loss function, $\Delta M(\Delta E)$, given in Ref.~\cite{Kavanagh:2020gcy}.\footnote{Data available at \href{https://github.com/bradkav/axion-miniclusters}{github.com/bradkav/axion-miniclusters}.} The functional dependence of the mass loss is in rough agreement with Ref.~\cite{Shen:2022ltx}, who instead performed idealised N-body simulations of an NFW halo interacting with a $1~M_{\odot}$ star. We notice mild disagreement only for $\Delta E/E_b\lesssim 10^{-2}$ and both models approach $\Delta M/M\sim 1$ as expected for very large energy injections $\Delta E/E_b\gg 1$ which can totally disrupt the minicluster, though these occur rarely. Reference~\cite{Shen:2022ltx} limits the treatment to NFW initial profiles only, while the modelling of Ref.~\cite{Kavanagh:2020gcy} can be extended to different profiles.

The initial energy of the minicluster before the energy injection is given by the sum of the kinetic energy ($1/2 M \sigma_{\rm mc}^2$) and binding energy ($E_b$):
\begin{equation}
    E_i =\left(\frac{\kappa}{2}-\beta\right) \frac{G M^2}{R}
\end{equation}
where $\kappa = 1.73,\,3.54$---for isolated and merged miniclusters respectively---parameterises the velocity dispersion as follows,
\begin{equation}\label{eq:sigmc}
    \sigma_{\rm mc}^2=\kappa\frac{GM}{R}\, .
\end{equation} 
Following energy conservation, the final minicluster energy after $\Delta E$ has been injected can be written as, 
\begin{equation}
    E_f = E_i+ (1-f_{\rm ej})\Delta E-E^{\rm unb}_i \, ,
\end{equation} 
$f_{\rm ej}(\Delta E)$ is the fraction of the imparted energy that becomes unbound and $E^{\rm unb}_i(\Delta E)$ the energy originally possessed by those axions that became unbound. Bound axions are those satisfying the integral cutoff in Eq.(\ref{eq:massloss}), and unbound axions the reverse. Like with the function $\Delta M(\Delta E)$ we adopt interpolations of $f_{\rm ej}(\Delta E)$ and $E^{\rm unb}_i(\Delta E)$ provided by Ref.~\cite{Kavanagh:2020gcy}. 

After each encounter, and before moving to the next one, the minicluster mass and radius are updated to,
\begin{equation}
    M\rightarrow M-\Delta M,\quad \quad R\rightarrow \left(\frac{\kappa}{2}-\beta\right)\frac{G(M-\Delta M)^2}{E_f} \, .
\end{equation}
We continue to perturb each minicluster through its $N_{\rm enc}$ encounters until either it is fully unbound ($E_f>0$) or we have reached the end of the orbit. Because many of the encounters occur very rapidly, we group together encounters in time bins of $\Delta t = t_{\rm MW}/10^4$, which is chosen to be shorter than the relaxation timescale $t_{\rm rel} \sim (G\rho_{\rm mc})^{-1/2}\sim \mathcal{O}(10\,{\rm Myr})$. This makes the procedure more computationally efficient whilst also ensuring that the relaxation to a new profile is only allowed to occur whenever there is sufficient time for it to occur (usually this is between disk crossings).

In addition to these successive mass-loss events, we also implement a second mass-loss procedure running in parallel to the stellar disruption in which the minicluster is continually stripped by the tidal field of the MW halo. The relevant formula for the rate of mass loss through this process is,
\begin{equation}
    \dot{M}=-\mathcal{A} \frac{M}{\tau_{\mathrm{dyn}}}\left(\frac{M}{M_{\rm MW}}\right)^\zeta \, ,
\end{equation}
where $\mathcal{A} = 1.34$ and $\zeta = 0.07$~\cite{Jiang:2014nsa}, $t_{\rm dyn} = 2.4$~Gyr is the dynamical time of the MW, and $M_{\rm MW} = 10^{12}~M_\odot$. The prescription for this is discussed in Appendix A of Ref.~\cite{Kavanagh:2020gcy} (see also Ref.~\cite{vandenBosch:2004zs}). Stripping by the MW halo contributes towards some of the early mass-loss experienced by the most massive miniclusters in our simulation, but is suppressed heavily by $(M/M_{\rm MW})^\zeta$ for smaller miniclusters. This also means it becomes increasingly negligible towards the end of the simulation.

\begin{figure*}
    \centering
    \includegraphics[width=\textwidth]{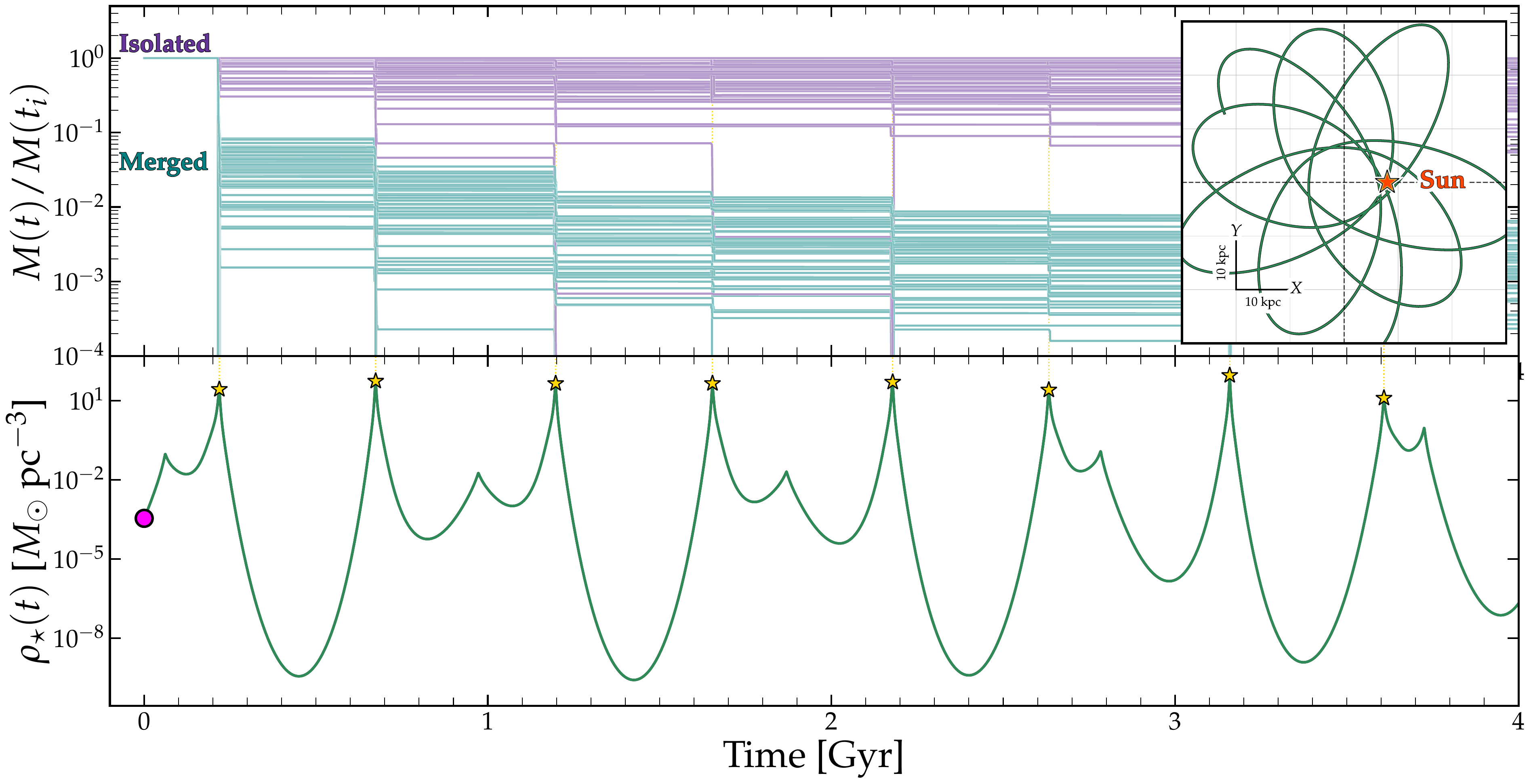}
    \caption{{\bf Upper panel:} the remaining masses of miniclusters for the first 4 Gyr of their evolution as they orbit the Milky Way, relative to their initial mass $M(t_i)$. The teal lines show 100 randomly sampled merged miniclusters (high-mass, NFW profiles), whereas purple is used for isolated miniclusters (low-mass, power-law profiles). All miniclusters are evolved along the same orbit. {\bf Lower panel:} The stellar density for the chosen orbit. The disk-crossing times are marked with yellow stars, which coincide with the events of major mass loss for both types of minicluster. {\bf Inset:} a projection of the trajectory of this chosen orbit in galactocentric $(X,Y)$ coordinates, where the Sun is located at (8,0) kpc.}
    \label{fig:evolution}
\end{figure*}
In Fig.~\ref{fig:evolution} we show the stellar density as a function of time, $\rho_{\star}(t)$ for a random orbit ending at the Solar position today $t = t_{\rm MW}$ (we zoom in on the initial 5 Gyr for visual clarity, although in our calculation we evolve each minicluster until today). The orbit has a mean galactocentric radius of $\sqrt{X^2+Y^2} = 22$~kpc and a mean inclination of $\sin^{-1}(Z/\sqrt{X^2+Y^2}) = 21^\circ$. In the upper panel, we show the gradual loss in mass as a function of time normalised to the minicluster's initial mass. The miniclusters are randomly sampled in mass from their mass function (see S.M.~\ref{app:sims}), but all follow the same orbit, an $X$-$Y$ projection of which is shown in the inset panel to the right. The mass loss exhibits a step-like behaviour for this orbit due to the exponential enhancement in the stellar density whenever the orbit crosses the disk plane $Z=0$. We see here clearly that the merged NFW-profile miniclusters lose a much larger fraction of their mass and over a much shorter timescale---particularly in the first disk-crossing. In contrast, many isolated power-law miniclusters survive with an $\mathcal{O}(1)$ fraction of their mass today.

\section{Impact of modelling uncertainties}\label{app:uncertainties}
To provide further intuition behind the results presented in the main text, in this section, we make some back-of-the-envelope estimates that explain why we should expect the DM density to be mostly replenished by the disruption of the most massive miniclusters. Doing this will also allow us to evaluate the possible effects (if any) that changing our modelling assumptions might have on our main result.

Let us start with a simplified description of the situation wherein the DM is composed entirely of merged miniclusters, which we assume are all disrupted in their entirety. This is already very close to our fiducial analysis since most of the mass is in the high-mass merged miniclusters anyway and, as we have seen (e.g.~in Fig.~\ref{fig:evolution}), the stellar disruption process causes them to lose the majority of their mass quite quickly. If we now further simplify the situation and assume all the resulting streams have the same volume $V_{\rm str}$ and the same mass $M$ then the expected number of streams overlapping at a single point is,
\begin{equation}
    \langle N_{\rm str}\rangle \approx N_{\rm mc} \frac{V_{\rm str}}{V} = \frac{\rho_{\rm DM} V_{\rm str}}{M} \, ,
\end{equation}
where in the second step we have rewritten the local galactic DM density we are required to saturate as \mbox{$\rho_{\rm DM} = N_{\rm mc} M/V$}. The expectation value for the density that we measure on Earth is then just the sum of all of the individual stream densities: $\langle \rho_{\rm obs} \rangle = \langle N_{\rm str}\rangle \rho_{\rm str} = N_{\rm str} M/V_{\rm str}$. Notice that if we substitute in $\langle N_{\rm str} \rangle$ from above we get back $\langle\rho_{\rm obs}\rangle = \rho_{\rm DM}$. We enforce this because when averaging the DM densities observed at many different points in a volume $V\gg V_{\rm str}$ should give us back DM density defined on that scale.

So the possible values of $\rho_{\rm obs}$ follow a binomial distribution with a mean of $\rho_{\rm DM}$. When $\langle N_{\rm str} \rangle$ is large, the central limit theorem dictates that the distribution of $\rho_{\rm obs}$ is a Gaussian centred on $\rho_{\rm DM}$. However in the opposite extreme, when $\langle N_{\rm str} \rangle \lesssim 1$ (for example if the miniclusters were not disrupted for whatever reason) the mean is still $\rho_{\rm DM}$, but only because the high probability to observe $\rho_{\rm obs} = 0$ is balanced against the very rare probability to observe a hugely enhanced density \mbox{$\rho_{\rm obs} = \rho_{\rm str}\gg \rho_{\rm DM}$} because the stream volumes would have to be very small to cause $N_{\rm str}\lesssim 1$. 

So for this reason, the pertinent result is not the expectation value of the density, because as a result of conserving the total mass of DM on large scales, this is always $\rho_{\rm DM}$ by construction\footnote{that is, in this simple example where our streams are objects of the same size and have constant internal densities.}. Instead, we must consider the distribution of $\rho_{\rm obs}$, which is dictated by $N_{\rm str}$. 

So to see which regime we are in, whilst also evaluating the effects of changing some of our assumed input parameters, let us develop a simple analytic estimate for the expected number of observed streams. Starting with the formula for the stream volume, parametrically this can be modelled as,
\begin{equation}
    V_{\rm str} \gtrsim \pi R^2 \sigma_{\rm mc} (t_{\rm MW}-t_{\rm enc}) \, ,
\end{equation}
where $4\sigma_{\rm mc} (t_{\rm MW}-t_{\rm enc})$ is roughly the stream length assuming a disruption happens (as we observe here) very quickly, i.e.~after only a few disk crossings. We use a $\gtrsim$ sign because the minicluster velocity dispersion gives us a lower limit on the stream's velocity dispersion. The former is expressed as $\sigma_{\rm mc}^2 = \kappa G M/R$, as in Eq.(\ref{eq:sigmc}) above. 

If we now plug in our formula for the radii of merged miniclusters $R(M) = R_{10} (M/10^{-10}\,M_\odot)^{1/3}$ (as in Eq.(\ref{eq:mergedR})), we see that the dependence on $M$ in the stream number $N_{\rm str} = \rho_{\rm DM} V_{\rm str}/M$ drops out entirely, giving us,
\begin{equation}
    N_{\rm str} = \frac{R_{10}^{3/2}}{(10^{-10} \, M_\odot)^{1/2}} 4\pi \sqrt{G \kappa} \rho_{\rm mc} (t_{\rm MW}-t_{\rm enc}) \gtrsim 200 \, ,
\end{equation}
where we assign $\rho_{\rm mc} = \rho_{\rm DM}(1-f_{\rm void}) \approx 0.01~M_\odot\,{\rm pc}^{-3}$ as the portion of the DM density that was originally bound inside miniclusters (i.e.~not in the minivoids). To evaluate this expression, we have taken the average mass-loss-weighted encounter time from our simulations of $t_{\rm enc}\sim 3$~Gyr, which gives us a number strikingly close to our fiducial numerical result. 

So referring back to our statements above about the distribution of $\rho_{\rm obs}$, the key point to highlight is that we are quite firmly in the regime $N_{\rm str}\gg 1$, especially because we have fixed the stream length $\sigma_{\rm mc}(t_{\rm MW} - t_{\rm enc})$ to be the \textit{smallest} that we expect it to grow to. We may account for additional heating of the axions to dispersions larger than $\sigma_{\rm mc}$, but will only act to increase this number, further reinforcing the fact that we expect there to be a large number of streams overlapping at our position.

\begin{figure}
    \centering
    \includegraphics[width=0.5\textwidth]{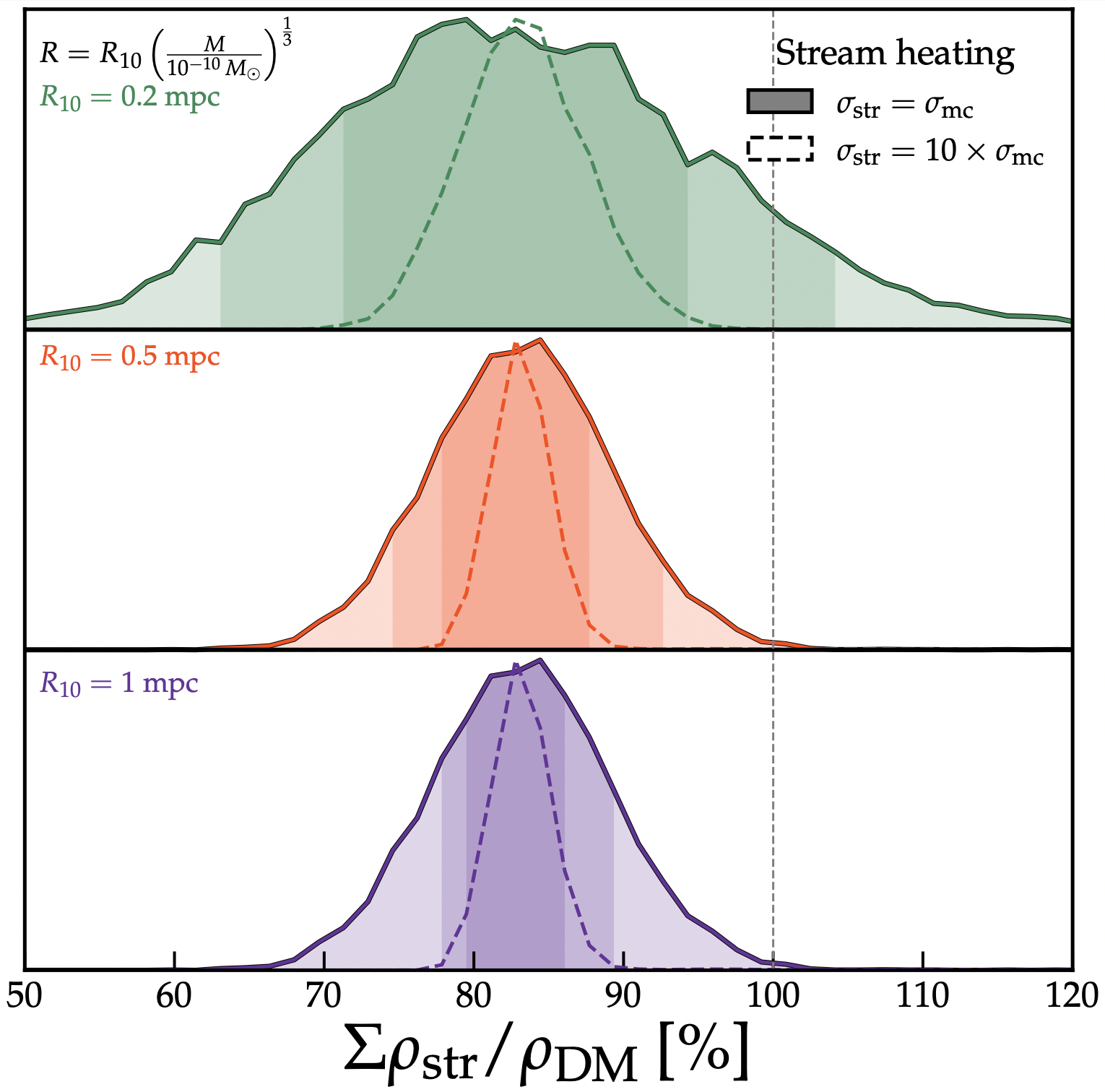}
    \caption{Independence of our final result on the minicluster mass-radius relation, parameterised by the radius scale $R_{10}$, and how much the stream's velocity dispersion is additionally heated from its initial value given by the minicluster's velocity dispersion. In all cases we see that the median value of observed density is roughly the same, with the primary role of these parameters being to change the expected number of overlapping streams, and hence the variance in the density, as opposed to its expectation value. The number of streams observed in the three examples ranges from $N_{\rm str} = 63,\,246,\,$ and $547$, from the top to bottom, and a factor of 10 larger for the $\sigma_{\rm str} = 10\sigma_{\rm mc}$ case.}
    \label{fig:MassRadiusRelation}
\end{figure}

This estimate also reveals a general insensitivity of the stream number to many model parameters. So we expect $N_{\rm str}\gg 1$ to remain the case if we allow the mass function, density profiles, mass-radius relation, and abundance of merged miniclusters to vary over any reasonable range. We can see this from our simple analytic estimate, but have confirmed this remains the case in our full simulation as well, which accounts for variable disruption probabilities as a function of $M$, varying disruption times as a function of the minicluster orbit, as well as the varying densities along the stream. In fact, the reason our numerical analysis leads to a median $\rho_{\rm obs}$ slightly smaller than $\rho_{\rm DM}$ is because of these additional details.

Nevertheless, our full numerical results are also generally insensitive to most of our more crude assumptions. To show a few explicit examples of factors that cause the result to change the most, Fig.~\ref{fig:MassRadiusRelation} shows multiple versions of our final result---the distribution in the sum of the stream densities, $\Sigma \rho_{\rm str}$---for multiple different choices for the initial sizes of the miniclusters $R_{10}$ (or equivalently, their concentration parameters, $c$), as well as for different stream heating factors beyond the most conservative case where $\sigma_{\rm str} = \sigma_{\rm mc}$. We see that even with these alternative choices, the median value of the density is mostly unchanged, and these quantities instead only affect the variance in the distribution (as expected because these quantities affect $\langle N_{\rm str}\rangle$). 

\section{Impact of minicluster halo substructure}\label{app:substructure}
As argued in the previous section, the conclusion that we expect the local DM density to be almost saturated after minicluster disruption can be considered reasonably robust, assuming our result that the merged miniclusters lose the majority of their mass holds up. However, it is this final issue that is really the only factor that could change our result in any substantial way. If the merged miniclusters are in fact more stable than the NFW halos that we have modelled them as, then it is possible they hold on to more of their mass and not spread it over such large volumes. One way in which this might not be captured in our treatment is if the merged miniclusters are not smooth NFW profiles but rather retain a degree of substructure from the isolated miniclusters that created them. We identify this as the primary remaining theoretical uncertainty that should be resolved with high-resolution (i.e. longer timescale) simulations that can evolve the minicluster halos to lower redshifts.

\begin{figure}
    \centering
    \includegraphics[width=0.5\columnwidth]{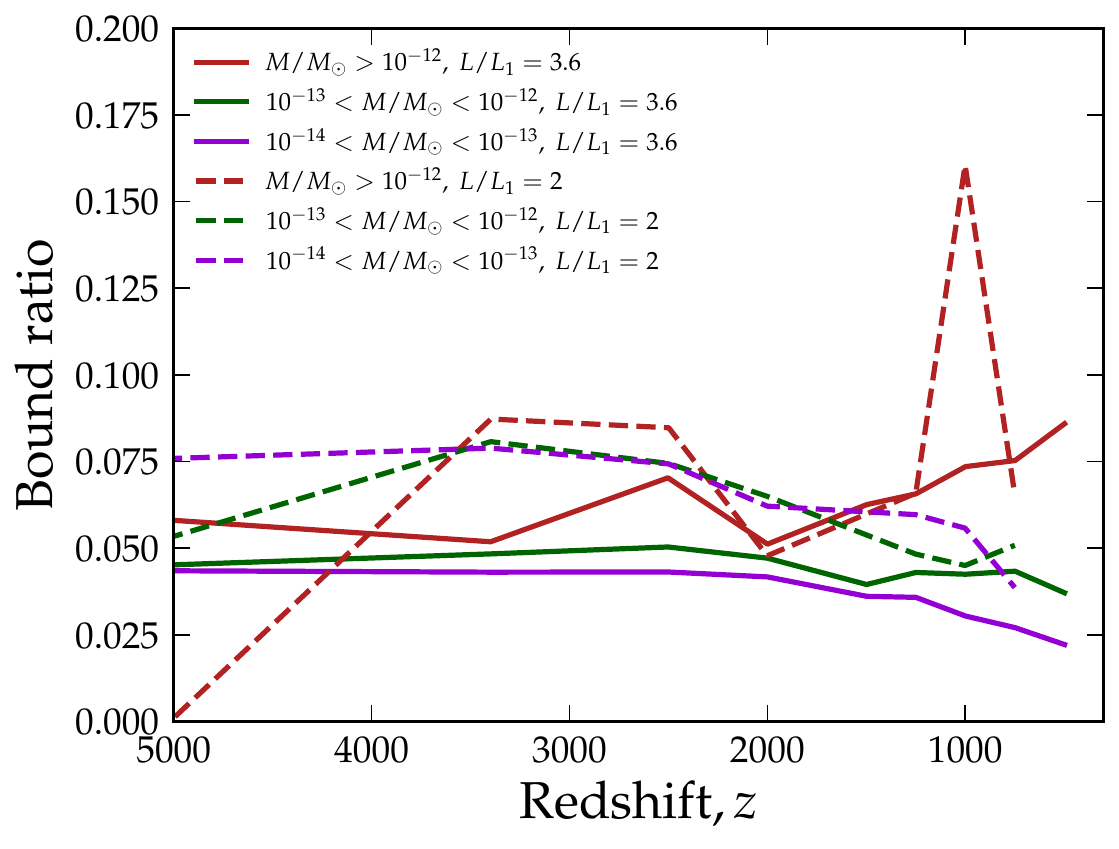}
    \caption{Bound fraction of axions within each minicluster halo (host), excluding the main virialised part, and as a function of the host mass.}
    \label{fig:isomer_hmf}
\end{figure}
So are our simulated minicluster halos actually just clusters of compact miniclusters? Minicluster halos are partially made by merging and accretion of isolated MCs. It is crucial for what follows to know the fraction of the mass of the MCH that is smoothly distributed and the fraction that it is still retained in compact MCs. This can be extracted from the numerical simulations to some extent using the \codo{subfind} algorithm implemented in the {gadget-4}~\cite{Springel:2020plp} code. From our recent N-body simulations (see Sec.~\ref{app:sims}) we calculate the fraction of axions contained in subhalos for each host (i.e., minicluster halo) identified by the \codo{friend-of-friends} (FOF) algorithm. In this calculation we however do not consider the main virialised halo (the first subhalo within the FOF group), since the latter will not contain itself additional substructure. As seen in Fig.~\ref{fig:isomer_hmf}, we find that only $\sim 10\%$ of the MCHs are composed of smaller virialised orbiting miniclusters. 
As expected, we confirm that high-mass MCHs have comparatively more substructure than low-mass MCHs. This lends further confidence that our modelling of these larger miniclusters as smooth NFW profiles can be considered reasonably accurate.


\end{document}